\documentclass[aps,prl,twocolumn,superscriptaddress]{revtex4}

\usepackage{graphicx}
\begin{document}

%

\preprint{Ver 3.0 \today}

\title{
Butterfly hysteresis loop at non-zero bias field in antiferromagnetic molecular rings: cooling by adiabatic
magnetization }

\author{
O. Waldmann } \email[Corresponding author.\\E-mail: ]{waldmann@physik.uni-erlangen.de}
\author{
R. Koch }
\author{
S. Schromm }
\author{
P. M\"uller } \affiliation{ Physikalisches Institut III, Universit\"at Erlangen-N\"urnberg, D-91058 Erlangen,
Germany }

\author{
I. Bernt }
\author{
R. W. Saalfrank } \affiliation{ Institut f\"ur Organische Chemie, Universit\"at Erlangen-N\"urnberg, D-91054
Erlangen, Germany }

\date{\today}

\begin{abstract}
At low temperatures, the magnetization of the molecular ferric wheel NaFe$_6$ exhibits a step at a critical
field $B_c$ due to a field-induced level-crossing. By means of high-field torque magnetometry we observed a
hysteretic behavior at the level-crossing with a characteristic butterfly shape which is analyzed in terms of a
dissipative two-level model. Several unusual features were found. The non-zero bias field of the level-crossing
suggests the possibility of cooling by adiabatic magnetization.
\end{abstract}


\maketitle

%

One of the most exciting recent discoveries in magnetism is that of magnetic hysteresis of purely molecular
origin in metal complexes like Mn$_{12}$ or Fe$_8$ \cite{Ses93,Gat94,Tho96,Fri96,San97,Gat00}. In these
large-spin molecules spin reversal is blocked at low temperatures due to an energy barrier. Recently, a
different type of molecular hysteresis has been observed in the molecule called V$_{15}$ \cite{Chi00}. Here, the
spin system is dynamically driven out of equilibrium leading to hysteresis loops with a characteristic butterfly
shape. The relevant time scale is on the order of seconds because of the phonon bottleneck effect. All in all,
V$_{15}$ is an experimental realization of a two-level system with dissipation \cite{Chi00}.

In this work, we investigated the molecular ferric wheel [NaFe$_6$L$_6$]Cl$\cdot$6CHCl$_3$ with L =
N(CH$_2$CH$_2$O)$_3$, or NaFe$_6$ in short \cite{Sal97}. Ferric wheels attracted much interest because of the
possibility of coherent quantum tunneling \cite{Taf94,Chi98}. In NaFe$_6$, the six iron(III) ions form a ring of
spin-5/2 centers with dominant antiferromagnetic nearest-neighbor Heisenberg interactions \cite{XFe6}. At zero
field the ground state is nonmagnetic ($S=0$), but application of a magnetic field leads to a level-crossing at
a critical field $B_c$ where the ground states changes abruptly to the first excited $S=1$, $M=-1$ level [inset
of Fig. 1(a)] \cite{Cor99,CsFe8}. Thus, from a general point of view, NaFe$_6$ represents a two-level system for
fields close to $B_c$. We will show that, for appropriate conditions, NaFe$_6$ exhibits a butterfly hysteresis
similar to that observed in V$_{15}$. This allows for important conclusions concerning fundamental physical
aspects of ferric wheels. Furthermore, we found several striking features of the spin reversal process,
demonstrating that the dissipative two-level system realized in NaFe$_6$ is actually a very distinctive one.

%

\begin{figure}[b]
\includegraphics{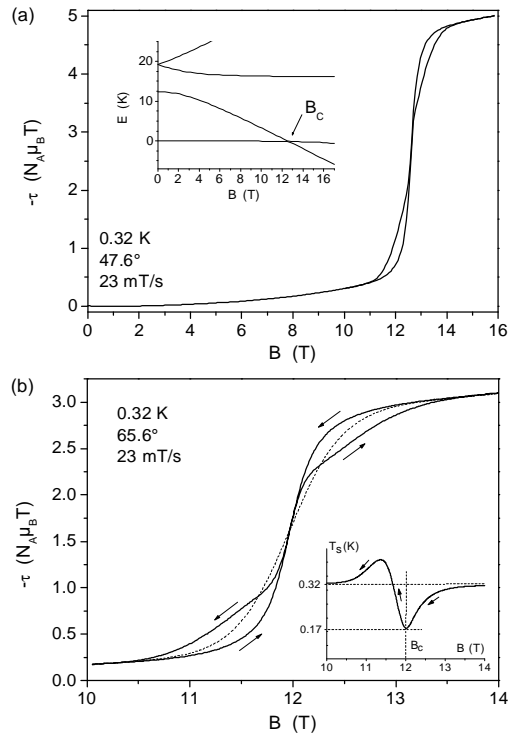}
\caption{Field dependence of the torque of a NaFe$_6$ single crystal demonstrating the butterfly hysteresis loop
as it typically appears for field sweep rates in the saturation regime (see text). The inset in panel (a)
schematically shows the field dependence of the low-lying energy states and the level-crossing at $B = B_c$
leading to the pronounced torque step. In panel (b), the dashed curve represents the measured equilibrium curve,
the inset sketches the behavior of the spin temperature $T_s$ when the field is swept from $B > B_c$.}
\end{figure}

The magnetic properties of single crystals of NaFe$_6$ were investigated by high-field torque-magnetometry. The
single crystals were prepared as reported \cite{Sal97}. They crystallize in the space group $R \bar 3$. The
cation [NaFe$_6$L$_6$]$^+$ exhibits $S_6$ molecular symmetry with the $S_6$ symmetry axis oriented perpendicular
to the plane of the cluster. The complex shows strict uniaxial magnetic behavior \cite{XFe6,CsFe8}. The uniaxial
axis coincides with the molecular $S_6$ symmetry axis. The torque was detected with homemade cantilever devices
micromachined from crystalline silicon \cite{CsFe8}. Resolution was typically 10$^{-11}$~Nm. For measurement, a
single crystal was selected by light microscopy in the mother liquor, was directly put from the solution into
Apiezon grease and mounted on the cantilever. The weight of the crystals investigated was typically 20~$\mu$g,
non-linearity was less than 1\%. The torquemeter was inserted into a 17~T cryomagnet with a vacuum loading
$^3$He insert. The angle between the crystal's uniaxial axis and magnetic field could be aligned {\it in situ}
with an accuracy of $0.3^\circ$. Calibration of the torque signal is accurate to $\pm$15\%. Temperature was
measured with a RuO$_2$ thick film resistor to within $\pm$30~mK.

%

A typical result for the field dependence of the torque at $T = 0.32\,$K is shown in Fig.~1. The negative sign
of the torque manifests the hard axis anisotropy of NaFe$_6$ \cite{XFe6}. The torque signal exhibits a steplike
field dependence at $B_c$ due to the abrupt switch of the ground state from the zero-field $S=0$ ground state to
the first excited $S=1$ state [inset of Fig.~1(a)]. This phenomenon is well known for ferric wheels
\cite{Cor99,XFe6,CsFe8}, but for the first time we also observed a pronounced hysteresis of the torque signal
for fields close to $B_c$, as shown in more detail in Fig.~1(b). The dependence of the butterfly curves on the
sweep rate of the field and on temperature is presented in Fig.~2. The field position of the curves, i.e. the
value of $B_c$, showed the expected angular dependence \cite{XFe6}, but otherwise we didn't observe an effect of
the angle.

\begin{figure}
\includegraphics{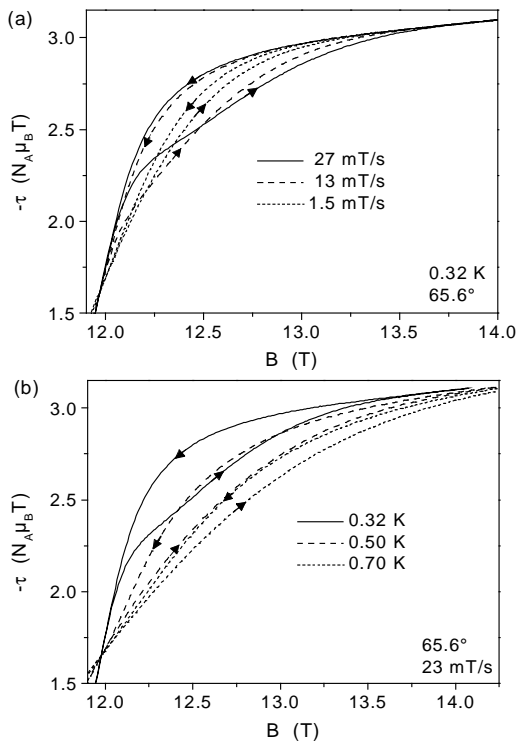}
\caption{Measured hysteresis loops for (a) three field sweep rates at $T$ = 0.32~K and (b) three temperatures at
a constant sweep rate of 23~mT/s. Only the branch with $B > B_c$ is plotted.}
\end{figure}

The underlying physics is quickly identified. Measurements performed with torque sensor and crystal immersed in
$^3$He exchange gas resulted in perfectly reversible torque curves. In this work, however, we used a vacuum
loading $^3$He insert and thermal contact between crystal and bath was established via the silicon cantilever.
In this configuration, thermal coupling should be weaker than with exchange gas. Obviously, the hysteretic
behavior shown in Fig.~1 reflects a non-equilibrium state due to limitations in heat flow indicating that
phonons play an important role in the process of changing the spin state at the level-crossing.

Similar butterfly hysteresis curves were observed recently for the molecular cluster V$_{15}$ and were
successfully described in terms of a phonon bottleneck scenario (PBS) \cite{Chi00,Compare}. The essence of the
PBS is that energy exchange between spin system and bath occurs via phonons \cite{Abr70}. The phonon bottleneck
arises because the number of spins is much larger than the number of available resonant phonon modes at low
temperatures. That is, the heat capacity of the spins $C_s$ is much larger than that of the phonons $C_p$ ($b =
C_s/C_p \gg 1$). Therefore, the phonon temperature $T_p$ follows that of the spin system $T_s$ very quickly (on
times $t_1 \approx b^{-1} t_s$) while energy transfer from the phonons to the bath is drastically delayed (time
scale $t_2 \approx b t_p$). Here, $t_s$ and $t_p$ denote the spin-phonon and phonon-bath relaxation times,
respectively. For the time scales of the present experiments, the spins and phonons may be regarded as a single
coupled system ($T_s = T_p$) which is weakly coupled to the bath ($T_s \neq T$).

The butterfly hysteresis curves reflect the behavior of the spin temperature in the PBS. Spin temperature and
torque are monotonously related: for a given magnetic field, $|\tau|$ is the larger the smaller $T_s$ if $B >
B_c$, and vice versa if $B < B_c$. Approaching the level-crossing from a starting field of e.g. $B > B_c$, the
spin temperature first decreases since the spin-phonon system cannot pick up heat quickly enough from the bath
[inset of Fig.~1(b)]. Thus, for sufficiently fast sweeping rates, the spin-phonon system is virtually decoupled
from the bath resulting in a (quasi) adiabatic demagnetization process. At $B \approx B_c$, the spin system
reaches a minimal temperature $T_{s,min}$. With further decreasing field, the spin system heats up (reverse of
adiabatic demagnetization) but overshoots the bath temperature since it now receives heat from the bath which
started to flow to reestablish equilibrium but reaches the spin system delayed due to the finite relaxation time
$t_2$.

The phonon bottleneck effect is obviously the underlying physical mechanism for the observed hysteresis in
NaFe$_6$, and in so far the situation is analogous to that of a conventional two-level system. However, there
are some particular characteristics to be noted, some of which are quite striking:

(1) With our setup we observed "heat-flow" hysteresis effects only for NaFe$_6$ (and related ferric wheels). We
have investigated quite a number of different molecular spin clusters, but for them the thermal coupling
provided by our silicon sensors was sufficient to achieve reversible torque curves. In this respect, the
observation of hysteresis in NaFe$_6$ is notable.

(2) A distinctive feature of the two-level system realized in NaFe$_6$ is that the two states involved belong to
different spin and spatial quantum numbers. In NaFe$_6$ the strong exchange limit is realized and the total spin
quantum number $S$ is a good (though not exact) quantum number \cite{XFe6,CsFe8}. Additionally, the cyclic
molecular symmetry implies a spin permutational symmetry \cite{Symmetrie} and states may be classified by the
wave vector $k$ \cite{CsFe8,Spindynamik}. The two states relevant in this work belong to $S = 0$, $k = \pi$ and
$S = 1$, $k = 0$, respectively, so that phonon absorption/emission is constrained by $|\Delta S| = 1$ and
$|\Delta k| = \pi$. To the best of our knowledge, the effects of these restrictions for the energy exchange
between spins and bath have not been investigated so far. But conceivably the number of available phonons may be
further reduced by these constraints. The $\Delta k$ selection rule may be of particular relevance since it
holds exactly as long as the molecular symmetry exhibits a $C_2$ symmetry axis.

(3) With a numerical analysis of the hysteresis curves based on the PBS, Chiorescu et al. evidenced a zero-field
splitting for the molecule V$_{15}$ \cite{Chi00}. In the case of ferric wheels, within a description in terms of
a spin Hamiltonian consisting of terms related to the electronic spins only, the level-crossing at $B = B_c$ is
a true crossing by virtue of translational symmetry \cite{CsFe8,Spindynamik}. However, interactions with e.g.
the nuclear spins may lead to a splitting of the levels at the crossing point \cite{Nor01,Mei01}. Their
identification would be of considerable importance for the issue of quantum coherence in antiferromagnetic rings
since they may act as sources of decoherence \cite{Mei01}. But so far, no clear evidence for a splitting of the
level-crossings, i.e. for anti-crossings, has been reported.

\begin{figure}
\includegraphics{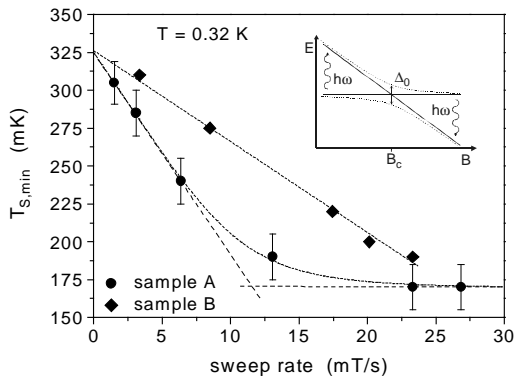}
\caption{Dependence of the minimal spin temperature $T_{s,min}$ on the field sweep rate for two samples. Dashed
lines are guides to the eye. The inset schematically presents the two states at the level-crossing and indicates
the splitting $\Delta_0$ discussed in the text.}
\end{figure}

We thus analyzed our data with the numerical procedure developed in Ref. [\onlinecite{Chi00}] as adapted to
torque measurements. Although we could generate curves which roughly resembled the experimental ones, we could
not reproduce the data for different temperatures and sweeping rates satisfactorily. Values for e.g. $\Delta_0$,
the splitting of the level-crossing at $B = B_c$ (inset of Fig.~3), showed unphysical scattering. This
indicates, that the PBS as outlined above is a too simple picture.

Therefore, we considered the dependence of the minimal spin temperature $T_{s,min}$ as function of the sweep
rate of the magnetic field. In conjunction with a model Hamiltonian, $T_{s,min}$ could be determined reliably
from the slope of the torque curves at $B = B_c$ as the curves are essentially reversible in this field regime.
In this way, $T_{s,min}$ could be estimated with a reproducibility of 5~mK, but the values are subject to a
systematic error of about $\pm$15~mK due to inaccuracies in the calibration of the torque signal. The results
for two samples are shown in Fig.~3. For sample $A$, $T_{s,min}$ first decreases linearly with increasing sweep
rate and then saturates at about 0.17~K. It should be noted, that the value of $T_{s,min}$ = 0.325~K
extrapolated to zero sweep rate agrees well with the bath temperature as measured with the RuO$_2$ sensor. In
the saturation regime, the spin system obviously behaves quasi adiabatically and it is thus very suggestive to
identify the saturation temperature as a level-splitting, i.e. $\Delta_0$ = 0.17~K.

We have also analyzed the broadening of the torque step as measured under reversible conditions for various
temperatures down to 0.4~K. This method provides an alternative way to estimate $\Delta_0$ \cite{Cor99,Nor01}.
However, even at the lowest temperatures the width of the step is in full accord with thermal broadening
resulting in an upper limit of 0.05~K for $\Delta_0$.

We have no explanation for this serious discrepancy. It suggests that the minimal spin temperature in the limit
of infinite sweep rates exceeds the thermodynamically determined splitting of the level-crossing. The PBS in the
present form obviously does not account for all the effects responsible for limiting $T_{s,min}$ and/or
broadening the torque step. It should be noted that the discrepancy cannot be explained by "extrinsic" effects
as e.g. crystal imperfections (twinning, mosaicity, etc.) \cite{Nor01}.

Incidentally, Fig.~3 demonstrates the role of the thermal coupling between crystal and bath. In case of sample
$B$ it is obviously better than for sample $A$ as the crossover is shifted to larger sweep rates.

\begin{figure}
\includegraphics{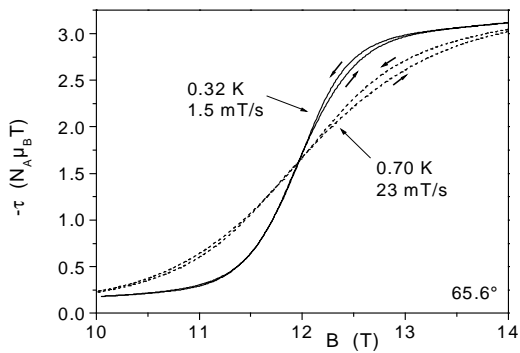}
\caption{Two example curves showing the asymmetry of the hysteresis loop around $B_c$ discussed in text.}
\end{figure}

(4) As mentioned already, $T_{s,min}$ as function of the sweep rate exhibits a crossover from a linear regime to
a saturation regime (Fig.~3). This is readily understood within the PBS. However, in the case of NaFe$_6$ the
crossover is accompanied by a striking change of the hysteretic behavior. In the saturation regime the curves
are symmetrical around $B_c$ [Fig.~1(b)], but in the linear regime the hysteresis becomes clearly asymmetric.
Two example curves are shown in Fig.~4. The asymmetry is more pronounced the slower the sweep rate or the higher
the (bath) temperature and was always found to be such that the hysteresis is larger for $B > B_c$ than for $B <
B_c$.

It seems as if energy exchange between spin system and bath is different for $B < B_c$ and $B > B_c$. This is
not easily explained. It is hardly conceivable that the thermal coupling between phonons and bath depends on
magnetic field. But then one has to conclude that the asymmetry is related to peculiarities of the spin-phonon
coupling in NaFe$_6$. However, the arguments presented above and in Refs.~[\onlinecite{Abr70}],
[\onlinecite{Chi00}] aim to show that spin-phonon coupling is so fast that in our experiments $T_s = T_p$ is
always maintained so that any details of the spin-phonon coupling are indiscernible! As demonstrated by the
symmetry of the curves for large sweep rates and under reversible conditions [Fig.~1(b)], an asymmetry of the
energy levels is not an issue.

(5) The most obvious distinctive feature of the NaFe$_6$ two-level system is the non-zero bias field of the
crossing, i.e. of $B_c > 0$. As shown in the above, the spin system cools down by approaching the critical field
$B_c$ as it is quasi adiabatically decoupled from the environment for large enough sweep rates. If one starts at
$B > B_c$ and decreases field, the situation corresponds to cooling by adiabatic demagnetization. However, if
one starts at $B < B_c$ (e.g. $B = 0$) and increases field, the reverse situation is realized, namely that of
cooling by adiabatic {\it magnetization}. Of course, in the present work no real adiabatic conditions were
realized - but the effect is clearly evident from Fig.~1.

%

In summary, we observed a butterfly hysteresis loop of the torque at the first ground state level-crossing in
the molecular ferric wheel NaFe$_6$. The results show that phonons are mostly responsible for transitions
between the two states near the level-crossing. This clarifies a recent proposal where it has been argued (as
hysteresis was not yet observed) that dipole-dipole coupling between nuclear and electronic spins plays an
important role \cite{Cor00}. The minimal spin temperature $T_{s,min}$ achieved during the hysteresis loop was
found to saturate for large sweep rates of the magnetic field - the first direct experimental evidence for
interactions of the electronic spins with the environment in ferric wheels. A detailed analysis of the processes
determining $T_{s,min}$ should provide deep insight into the conditions for quantum coherence in ferric wheels.
The present work suggests a novel experimental route to measure them.

For slow sweep rates the hysteresis loop exhibited a striking asymmetry which is unexplainable within a standard
phonon bottleneck scenario. One may only speculate about the role played by the unusual quantum numbers of the
states involved in the level-crossing. It also has been demonstrated that the NaFe$_6$ system allows for cooling
by adiabatic magnetization, i.e. for the opposite of what is generally known. At the level-crossing the ferric
wheel NaFe$_6$ represents a dissipative two-level system - but with quite unusual properties sheding new light
on a model of which one thought to know all.

%
\begin{acknowledgments}
OW would like to thank I. Chiorescu for fruitful comments concerning the numerical simulation of the phonon
bottleneck effect. This work was supported by the Deutsche Forschungsgemeinschaft, the Sonderforschungsbereich
583, and the Bayerisches Langzeitprogramm "Neue Werkstoffe".
\end{acknowledgments}

%

%

\begin{references}
\bibitem{Compare}
Please note that in Ref. \onlinecite{Chi00} magnetization was measured while we used torque. However, for
NaFe$_6$, the gap between the two lowest levels depend on field and angle approximately as $\Delta(B,\phi) =
\Delta(B-B_c(\phi ))$ for fields close to $B_c$. In this approximation, torque is directly proportional to
magnetization.
\end{references}
\end{document}